\documentclass[twocolumn,showpacs,preprintnumbers,amsmath,amssymb]{revtex4}
\usepackage{graphicx}
\usepackage{dcolumn}
\usepackage{bm}

\begin{document}


\title{New Limits on Naturally Occurring Electron Capture of $^{123}$Te}

\author{A.~Alessandrello, C.~Arnaboldi, C.~Brofferio, S.~Capelli, O.~Cremonesi, E.~Fiorini, A.~Nucciotti, M.~Pavan, G.~Pessina, S.~Pirro, E.~Previtali, M.~Sisti, M.~Vanzini and L.~Zanotti}
\affiliation{
Dipartimento di Fisica dell'Universit\'a di Milano-Bicocca\\
e Sezione di Milano dell'INFN\\
I-20133 Milano, Italy.}
\author{A.~Giuliani e M.~Pedretti}
\affiliation{
Dipartimento di Scienze Chimiche, Fisiche e Matematiche  dell'Universit\'a d'Insubria\\
e Sezione di Milano dell'INFN\\
 I-20133 Milano, Italy.
}
\author{C.~Bucci}
\author{C.~Pobes}
\altaffiliation{CEE fellow  in the Network on Cryogenic Detectors, under contract FMRXCT980167}
\affiliation{
Laboratori Nazionali del Gran Sasso\\
I-67010, Assergi (L'Aquila), Italy.
}
\

\begin{abstract}
Electron capture of $^{123}$Te from the K shell has been investigated in a new underground search with an array of  340 g TeO$_2$ thermal detectors. We find that some previous indication of this decay could be attributed to E.C. of $^{121}$Te resulting from neutron activation of natural Tellurium. There is therefore so far {\it no evidence} for  E.C. of $^{123}$Te from the K shell with a 90\% c.l. lower limit $t_{1/2}^K>5 \times 10^{19}$ years on the half lifetime. Taking into account the predicted K E.C. branching ratio, the corresponding lower limit on the $^{123}$Te EC half lifetime is $t_{1/2}>9.2 \times 10^{16}$, which can be theoretically interpreted only on the basis of a strong suppression of the nuclear matrix elements. A complementary analysis based on the expected fraction of E.C. accompanied by internal bremsstrahlung is discussed.
\end{abstract}

			      
\maketitle

\section{Introduction}
Single beta decay or electron capture is expected to occur in nine isobaric doublets or triplets existing in nature \cite{audi54}. It has been actually detected in seven of them \cite{endt90,burr90,siev91,lytt91,ales94,geor97,blac87,tani00,fire91,ashk93,ales99,fonta00}, while no evidence has been found for beta decay of the 9$^-$ excited state of $^{180}$Ta \cite{brow94}. More complex is the situation \cite{ohya93,fire98,watt62,ales96} for the second forbidden unique electron capture (EC) of  $^{123}$Te to the ground state of $^{123}$Sb, with a transition energy of  53.3 $\pm$ 0.2 keV (Q$_{EC}$).\par 
Contradictory experimental results have been in fact presented~\cite{watt62,ales96}. Very low rates have been, on the other hand,  predicted theoretically on the basis of a strong suppression of the nuclear matrix elements due to a cancellation between particle-particle and particle-hole correlations. 
As suggested By M.~Bianchetti et.~al.~\cite{bian97} it is expected that these cancellations can lead to a suppression of the rate by up to six orders of magnitude. In a more recent and detailed paper, O.~Civitarese and J.~Suhonen~\cite{civi01} note that a suppression of the nuclear matrix element of the second-forbidden EC transition of this nucleus is a very severe test of the nuclear models used to calculate rare electroweak decays.
One of the claimed experimental evidences of the decay~\cite{watt62} is in contradiction with this thoretical model. 
A precise measurement of the $^{123}$Te decay rate can therefore shed light both on the experimental situation and on the mechanisms which are at the basis of the above mentioned cancellation. 
EC processes, can be observed through the measurement of the inner bremsstrahlung (IB) photons accompanying a generally small fraction of decays, or through the atomic deexcitation cascade (X-rays and/or Auger electrons) following the decay. 
Individual atomic transitions (single X-rays or Auger electrons) can be observed using a detector in direct contact, but external to the decay source. A single line corresponding to the binding energy of the captured electron (sum of all atomic transition energies) is on the contrary expected for a pure calorimetric approach, in which the source is also the detector (as is the case for a low temperature thermal detector~\cite{boot96}).
\begin{figure}[!bth]
\includegraphics[height=.45\textheight]{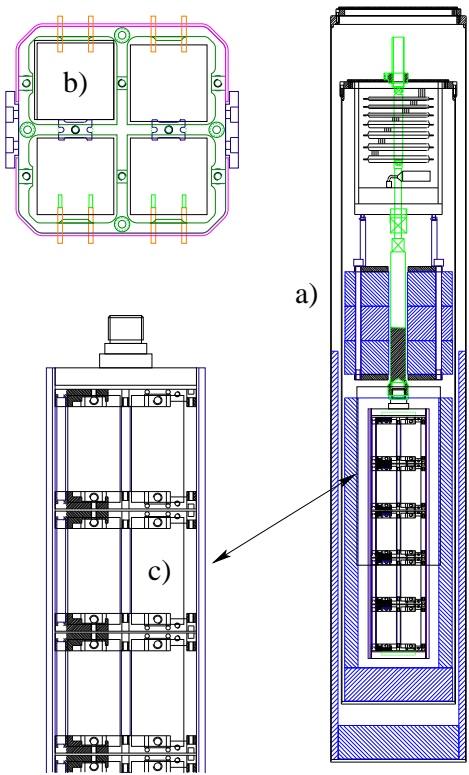}
\caption{\label{fig:setup} Cryogenic setup used in the new measurement (a). Details of the inner Roman lead shield are apparent. Top (b) and side (c) view of the 20 TeO$_2$ crystal array are also shown.}
\end{figure}
Unfortunately, atomic deexcitation cascades can be induced also by the interaction between the environmental radiation and the source (e.g. photoelectric or Compton effect). 
In this case, however, the involved atomic levels are those of the parent atom, while in a genuine EC decay they correspond to those of the daughter one (Table \ref{tab:dede}).

\section{Previous experimental results}
Evidence for the K EC decay of $^{123}$Te was obtained by Watt and Glover~\cite{watt62} with a lifetime $t_{1/2}^K=(1.24 \pm 0.10)  \times 10^{13}$ years, using a proportional counter. Such a value is still reported in the Nuclear Tables~\cite{fire98}.
Only the X-rays escaping from the Te source (anode wires) could be recorded in such an experiment. Furthermore, due to the insufficient energy resolution of the proportional counter, there was no possibility to discriminate between the Sb X-ray line at 26.1 keV, distinctive of Te EC decay, and the 27.3 Te X-ray line due to the excitation of the tellurium source by cosmic rays and radioactivity. The inclusion of a non negligible background contribution (the experiment was carried out at sea level) could explain therefore why the authors obtained a so large rate for this process.\par
This result was contradicted by a a previous cryogenic experiment (Run 0) carried out underground by our group using a low activity setup consisting of four thermal 340 g TeO$_2$ detectors. A description of this set-up and the operation of the anticoincidence is reported in ref. ~\cite{ales96}. In addition to the almost complete elimination of the external background due to cosmic rays, special care was devoted to the reduction of background from environmenthal radioactivity. Since the backround due to internal contamination of the crystals was negligible, the surface of the crystals and the wall of the copper frame facing them was
treated to avoid contribution from surface contamination. External background was strongly reduced with layers of lead of minimum thickness of 20 cm.

Thanks to the adopted calorimetric approach and the good energy resolution of the detectors, we could clearly distinguish two peaks in the spectrum recorded at low energy: a peak at 27.3 keV corresponding to the energy of Te K$_\alpha$ X-rays (produced by the interaction of radiation with nearby Te detectors), and a peak at 30.5 keV, corresponding to the total energy released by Te K EC to Sb. The different origin of the two peaks was moreover demonstrated by the comparison of the spectra collected requiring or not an anticoincidence between the four TeO$_2$ detectors~\cite{ales96}: the 27.3 keV line in fact disappeared in the anticoincidence spectra. 
By attributing the 30.5 keV peak to K EC of $^{123}$Te we obtained an evidence for this process and quoted a lifetime $t_{1/2}^K=(2.4 \pm 0.9) \times 10^{19}$ years, six orders of magnitude higher than in the experiment by Watt and Glover. 
The main drawback of our previous experiment result was the limited statistics.
In addition we were worried that the 30.5 keV signal could be due to the activation of the $^{120}$Te isotope (0.908~\% abundance) by environmental neutrons. This isotope, despite its low abundance can in fact lead  to a substantial production of $^{121}$Te and $^{121m}$Te with cross sections of 0.3 and 2 b respectively. These isomers decay by EC with lifetimes of 16.8 and 154 days, respectively, yielding the same signal in the detector as the one expected for the EC decay of $^{123}$Te. The expected neutron activation in the underground laboratory is negligible, since thermal neutrons are suppressed by a factor of ~10$^4$ with respect to sea level~\cite{bett01}. However, $^{121}$Te and $^{121m}$Te nuclei could have been produced when the detectors were outside the laboratory. They could have persisted in the detectors during the measurement, since the run started underground immediately after the crystal installation.

\section{Structure and results of the improved set-up}
These considerations lead us to perform a new improved underground measurement. A larger cryogenic setup consisting of an array of twenty 340 g crystals of natural TeO$_2$~\cite{ales98} was operated (Fig.\ref{fig:setup}).\par
\begin{figure}[!th]
\includegraphics[height=.27\textheight]{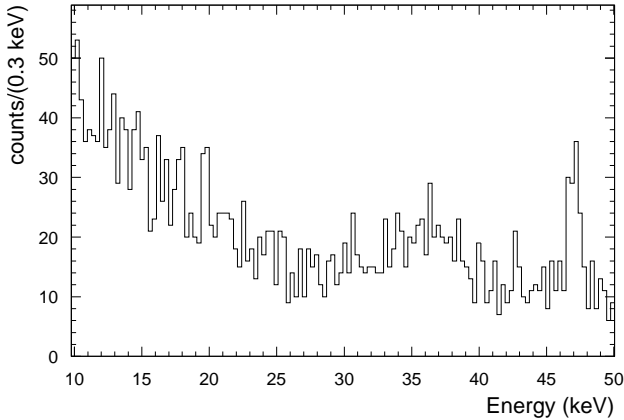}
\caption{\label{fig:20old} Low energy spectrum obtained in the run carried out two years after the crystals were stored underground where they were shielded against cosmic ray neutrons (Run 1 in the text).The peak in the spectrum, corresponding to the 46.6 line of $^{210}$Pb has a width of 1 keV FWHM.}
\end{figure}
Specifically relevant for the search reported here is the reduced background in the low energy region, mainly due to the addition of a low activity Roman lead shield close to the detectors (10 cm above and below the detectors and of 1 cm around them). Due to the absence of $^{210}$Pb\cite{ales98b}, this material is very effective in reducing the counting rate at low energy.\par
This new experiment consists in two separate runs. The former (Run 1) started two years after the crystals were stored underground, in order to allow a substantial decay of the $^{121}$Te and $^{121m}$Te nuclei produced during their preparation. 
The spectrum corresponding to 724.88 hours~$\times$~kg of effective running time is shown in Fig.~\ref{fig:20old}. The excellent resolution achieved (1 keV FWHM) clearly indicates the stability of the experiment. With respect to our previous search the background is reduced by about an order of magnitude, due to better cleaning of the surfaces and introduction of the internal lead shields. This explains the disappearance of the 27 keV X-ray peak, due to background excitation of Tellurium. Moreover, the peak at 39.9 keV due to $^{212}$Bi observed in the spectrum of the previous experiment, is no more apparent due to the effective reduction of Th contaminations in the new setup. The peak at 46.5 keV is due to surface activity of $^{210}$Pb, a common surface contaminant in all these experiments \cite{heus95}. The same contamination is also responsible of the bump at $\sim$~35 keV. As confirmed by Monte Carlo simulations this bump is the consequence of the ensemble of processes ($\gamma$ absorbtion and atomic rearrangement) following the $\beta$ decay of $^{210}$Pb to the 46.5 keV excited state of $^{210}$Bi.\par

\begin{figure}[!bth]
\includegraphics[height=.27\textheight]{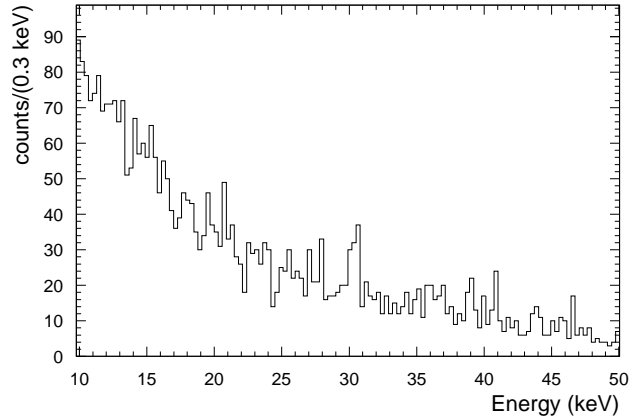}
\caption{\label{fig:20new} Spectrum obtained in a run initiated few days after the crystals were exposed for two months to cosmic ray neutrons (Run 2 in the text).}
\end{figure}

Unlike what had been indicated in the previous experiment, there is now no clear evidence of the peak at 30.5 keV, where the observed counts are 17 $\pm$ 12. 
In order to assess the origin of this disagreement we carried out a second run taking advantage of a long measurement stop during which the experimental setup was completely unmounted and rebuilt with the aim to further improve the background level. In particular, all crystals were brought outside the underground laboratory for a surface treatment aiming at a reduction of their surface radioactive contamination. They remained therefore exposed to environmental neutrons for a period of about two months. After this operation they were again installed underground and a second run (Run 2) totalling 259.59 hours~$\times$~kg took place. The low energy spectrum reported in Fig.\ref{fig:20new} clearly shows a peak at 30.5 keV which could now be attributed to the K EC of $^{121}$Te isomers produced by cosmic ray neutrons outside the tunnel and not to K EC of $^{123}$Te.\par

Taking into account:

\begin{itemize}
\item the absence of a clear signal in our present data (Run 1);
\item the short period during which the TeO$_2$ crystals were stored underground before the start of our previous experiment (Run 0);
\item the evidence for a neutron activation contribution to EC signals (30.5 keV line) in crystals exposed to sea level neutron flux and not stored underground for a long enough period (Run 2);
\end{itemize}

we conclude that {\it there is so far no evidence for K electron capture of $^{123}$Te}. 
In particular, by applying a maximum likelihood analysis to the data collected during Run 1 we can set a 90~\% C.L. lower limit $t_{1/2}^K \equiv \frac{t_{1/2}}{BR^K} > 5 \times 10^{19}$ years on the half lifetime for K EC of $^{123}$Te. Assuming a branching ratio for K capture $ BR^K=1.83\times10^{-3}$~\cite{bian97,bamb77}, the inclusive limit for EC of $^{123}$Te is $t_{1/2}>9.2 \times 10^{16}$ years (90~\% C.L.). More theoretical efforts are needed to explain this strong suppression with respect to similar $\beta$ processes occurring in nature.\par

\begin{table*}[!bth]
\caption{\label{tab:dede} Summary of all possible atomic readjustment processes and of their detection modes with conventional radiation detectors and true calorimeters. Only the K shell case is reported. The same is obviously true also for the L shell electron capture whose peak, due to its lower energy, cannot however be detected in our experiment.
}
\begin{ruledtabular}

\begin{tabular}{lcc}
Atomic readjustment origin&
True calorimetric approach& 
External radiation detector\\

\hline
&&\\
{\bf K shell Electron Capture}&
Sum of the energies of &
Independent atomic transitions\\

(Atomic levels of the daughter atom: Sb)&
of all atomic transitions &
(K$_\alpha^{Daughter}$, K$_\beta^{Daughter}$, etc.)\\
&(K electron binding energy)&\\

\hline
&{\bf Source~$\ne$~detector}&\\

{\bf Radiation induced processes} &
Independent atomic transitions&
Independent atomic transitions\\

{\bf in the source} &
(K$_\alpha^{Parent}$, K$_\beta^{Parent}$, etc.) &
(K$_\alpha^{Parent}$, K$_\beta^{Parent}$, etc.)\\

(Atomic levels of the parent atom: Te)&{\bf Source~=~detector}&\\
&K electron binding energy&\\
&+ radiation deposited energy&\\
\end{tabular}

\end{ruledtabular}
\end{table*}
A limit for the overall decay rate of $^{123}$Te can also be evaluated on the basis of the predicted probability of internal bremmstrahlung accompanying electron capture from any atomic shell.\par
A continuous spectrum spanning the energy region up to Q$_{EC}$ minus the binding energy of the captured electron (B$_{el}$) is expected for a conventional radiation detector. Since however the atomic deexcitation energy is fully detected in the case of a calorimetric approach, the spanned energy region starts at B$_{el}$, up to Q$_{EC}$. The probability to observe bremsstrahlung photons of energy between 25 and 45 keV in our ``calorimeter'' is $\sim$ 0.14~\%~\cite{tabi}. This figure includes both the branching ratio and the detector efficiency.
A total of (1067~$\pm$~33) counts was recorded in this energy region in our experiment, allowing us to set, conservatively, an inclusive lower limit of $t_{1/2}>2 \times 10^{15}$ years (90~\% C.L.) for electron capture of $^{123}$Te in any decay channel.
Though statistically poorer such a result provides a complementary determination of the half lifetime for this process, independent of the decay channel, thus confirming the strong rate suppression observed for the K EC mode.\par
%

Thanks are due to the Laboratori Nazionali del Gran Sasso for generous hospitality. We gladly acknowledge the help of P.~Gorla, S.~Parmeggiano and L.~Tatananni in various stages of this experiment. Special thanks are due to M.~Perego for assembling the electronic read-out and to A.~Rotilio for the realization of mechanical parts. 
	Very fruitful discussions with the theoretical nuclear physics groups of the University of Milano and particularly with P.F.~Bortignon, R.A.~Broglia and G.~Col\`o are also gratefully acknowledged. 
\\
\\
This experiment has been supported in part by the Commission of European Communities under contract FMRXCT980167.

\end{document}